\documentclass[aps, prl, reprint, twocolumn, superscriptaddress]{revtex4-1}

\usepackage{multirow}
\usepackage{graphicx}
\usepackage{longtable}
\usepackage[utf8]{inputenc}
\usepackage[T1]{fontenc}
\usepackage{epstopdf}
\usepackage{textcomp}
\usepackage[cmyk,dvipsnames]{xcolor} 
\usepackage{amsbsy}
\usepackage{amsmath}
\usepackage{amssymb}
\usepackage{amsfonts}
\usepackage{dsfont}
\usepackage{mathtools}
\usepackage{braket}
\usepackage{array}
\usepackage{gensymb}
\usepackage{placeins}
\usepackage{dashrule}
\usepackage{xcolor}
\usepackage{booktabs}
\usepackage{hyperref}
\usepackage{float}
\usepackage{braket}
\usepackage{ulem}

\usepackage[draft]{pgf}
\usepackage{pgfplotstable}
\usepackage{booktabs}
\usepackage[inline]{enumitem}

\def\pgfglobalkeys#1{\begingroup \ifnum\the\globaldefs>0\relax \else \globaldefs=1\fi \pgfkeys{#1}\endgroup}

\pgfplotstableset{display column names/.code={
        \foreach[count=\n from 0] \x in {#1} {
            \pgfglobalkeys{/pgfplots/table/display columns/\n/.estyle={column name=\x}}
        }
    }
}

\newcommand{\hess}{H}
\newcommand{\carthess}{\bar{\hess}}

\newcommand{\vecext}[1]{#1}
\renewcommand{\vec}[1]{\mathbf{#1}}

\makeatletter
  \def\my@tag@font{\normalsize}
  \def\maketag@@@#1{\hbox{\m@th\normalfont\my@tag@font#1}}
  \let\amsmath@eqref\eqref
  \renewcommand\eqref[1]{{\let\my@tag@font\relax\amsmath@eqref{#1}}}
\makeatother

\allowdisplaybreaks

\begin{document}

\title{Duplication, collapse and escape of magnetic skyrmions revealed using a systematic saddle point search method}

\newcommand{\fz}{Peter Gr\"unberg Institut and Institute for Advanced Simulation, Forschungszentrum J\"ulich and JARA, 52425 J\"ulich, Germany}
\newcommand{\iceland}{Science Institute and Faculty of Physical Sciences, University of Iceland, VR-III, 107 Reykjav\'{i}k, Iceland}
\newcommand{\rwth}{RWTH Aachen University, D-52056 Aachen, Germany}
\newcommand{\ITMO}{ITMO University, 197101, St. Petersburg, Russia}
\newcommand{\SPBSU}{Dpt.~of Physics, St. Petersburg State University, St. Petersburg 198504, Russia}
\newcommand{\Aalto}{Dpt.~of Applied Physics, Aalto University, FIN-00076 Espoo, Finland}

\author{Gideon P. M\"uller}
	\affiliation{\iceland}
 	\affiliation{\fz}
        \affiliation{\rwth}
\author{Pavel F. Bessarab}
    \affiliation{\iceland}
	\affiliation{\ITMO}
\author{Sergei M. Vlasov}
	\affiliation{\iceland}
	\affiliation{\ITMO}
\author{Fabian Lux}
	\affiliation{\fz}
	\affiliation{\rwth}
\author{Nikolai S. Kiselev}
	\affiliation{\fz}
\author{Stefan Bl\"ugel}
 	\affiliation{\fz}
\author{Valery M. Uzdin}
	\affiliation{\ITMO}    
	\affiliation{\SPBSU}
\author{Hannes J\'onsson}
 	\email[]{hj@hi.is}
	\affiliation{\iceland}
	\affiliation{\Aalto}


\begin{abstract}
Various transitions that a magnetic skyrmion can undergo are found in calculations using a method for climbing up the energy surface and converging onto first order saddle points. In addition to collapse and escape through a boundary, the method identifies a transition where the skyrmion divides and forms two skyrmions. The activation energy for this duplication process can be similar to that of collapse and escape. A tilting of the external magnetic field for a certain time interval is found to induce the duplication process in a dynamical simulation. Such a process could turn out to be an important avenue for the creation of skyrmions in future magnetic devices.
\end{abstract}



\maketitle


Localized, non-collinear magnetic states are receiving a great deal of attention, where skyrmions have come under special focus.
Along with interesting transport properties, skyrmions exhibit particle-like behaviour and carry a topological charge enhancing their stability with respect to uniform ferromagnetic background.
In addition to the interest in their intriguing properties, they have been suggested as a basis for technological applications e.g. data storage or even data processing devices \cite{Kiselev11,Fert17}.
A racetrack design of a memory device has been outlined where a spin polarized current drives a chain of skyrmions past a reading device \cite{Fert13,Muller17}.
The effect of temperature and external magnetic field on the stability of the skyrmions need to be studied, as well as
ways to generate and manipulate them. The effect of defects is also an important consideration \cite{Muller15,Hanneken16}. 
Two mechanisms for the annihilation of skyrmions have been characterized by theoretical calculations of atomic scale systems: Collapse of a skyrmion to form ferromagnetic state \cite{bessarab_method_2015,Lobanov16,Stosic17,Cortes-Ortuno17,Uzdin18} and escape of a skyrmion through the boundary of the magnetic domain \cite{bessarab_2018,Stosic17,Cortes-Ortuno17,Uzdin18}. 
The effect of a non-magnetic impurity has also been calculated \cite{Uzdin18}. 
By using harmonic transition state theory for magnetic systems \cite{bessarab_harmonic_2012,bessarab_potential_2013}, the lifetime of skyrmions has been estimated \cite{bessarab_2018,Uzdin18}.
Parameter values obtained from density functional theory \cite{Malottki_2018} are found to give
results that are consistent with experimental observations \cite{Romming_2013,Kubetzka_2017}.  
The challenge is to design materials where magnetic skyrmions are small enough while being sufficiently stable at ambient temperature, and to develop methods for manipulating them. 

Theoretical calculations can help accelerate this development 
by identifying the various possible transformations that a skyrmion can undergo at a finite temperature on a laboratory time scale. 
This can be achieved by the use of rate theory where the major challenge is to find the relevant transition mechanisms. If the final state of a transition is specified, in addition to the initial state, the geodesic nudged elastic band (GNEB) method \cite{bessarab_method_2015,Bessarab_2017} 
can be used to find the minimum energy path of the transition and, thereby, the activation energy which is the highest rise in energy along the path.
However, the final states of possible transitions are not always known. 
Another category of methods 
for identifying possible transition mechanisms where the final states are not specified, only the initial state, 
has turned out to be highly valuable
in a different context, namely in studies of atomic rearrangements such as chemical reactions and diffusion events 
\cite{Peters17,Jonsson11}.  
Unexpected transition mechanisms have in many cases turned out to be preferred over mechanisms that seem {\it a priori} most likely
\cite{Johannesson01}.

Here, we describe a method that can be applied to identify transition mechanisms in magnetic systems without specifying final states. 
It represents an adaptation of a method that has been used extensively in studies of atomic rearrangements. 
A complication arises from the fact that magnetic systems are characterized by the 
orientation of the magnetic moments while the length of the magnetic moments is either fixed or obtained from 
self-consistency calculations \cite{Bessarab_2014}.  
The configuration space is therefore curved.  
The method is first described briefly, with more detailed information in Supplemental Material \cite{SupportingInfo}, and then an application to a magnetic skyrmion is presented
where, in addition to collapse and escape, the method gives a mechanism and activation energy 
that has not been reported before: Duplication of a skyrmion.
Finally, a dynamical simulation is described where a time dependent external field is used to induce such an event. 


Within harmonic transition state theory \cite{Peters17,bessarab_harmonic_2012,bessarab_potential_2013}, 
the mechanism and rate of a thermally induced transition is characterized by the first order saddle point \cite{FirstOrderSP}
representing the bottleneck for the transition. 
Given an initial state corresponding to a local energy minimum, the various possible transitions the system may undergo can be identified by climbing up the energy surface and converging on the various first order saddle points on the energy ridge surrounding the minimum. 
A version of this approach, referred to as minimum mode following 
\cite{Henkelman99,Gutierrez17}, is illustrated for a single-spin test system in Fig. 1. 
The method is based on the evaluation of 
eigenvalues and corresponding eigenvectors of the Hessian, $\hess$.
First, the region near the minimum, the convex region where all eigenvalues of the Hessian are positive, is escaped by following, for example, the gradient of the energy. Alternatively, a random vector or one of the eigenvectors of $\hess$ can be followed to escape the convex region. Once an eigenvalue turns negative, an effective force 
%
%
\begin{equation}
	\vecext{F}^\mathrm{eff} = \vecext{F} - 2 (\hat{\vecext{\lambda}}\cdot\vecext{F})~\hat{\vecext{\lambda}} ,
    \label{eq: effective force}
\end{equation}
is followed, where $\vecext{F}=-\vecext{\nabla} \mathcal{H}$ is the negative gradient of the energy
and $\hat{\vecext{\lambda}}$ is the normalized eigenvector corresponding to the negative eigenvalue. Note that these vectors are $3N$-dimensional for a system with N spins.
As the system is displaced in the direction of the effective force,
it moves to higher energy along $\hat{\vecext{\lambda}}$ but to lower energy along the orthogonal degrees of freedom. 
Eventually, this brings the system to a first order saddle point on the energy surface. 
The final state of the transition can be obtained by a slight displacement further along $\hat{\vecext{\lambda}}$, 
followed by energy minimization.

\begin{figure}[!h]    
		\includegraphics[width=0.8\linewidth]{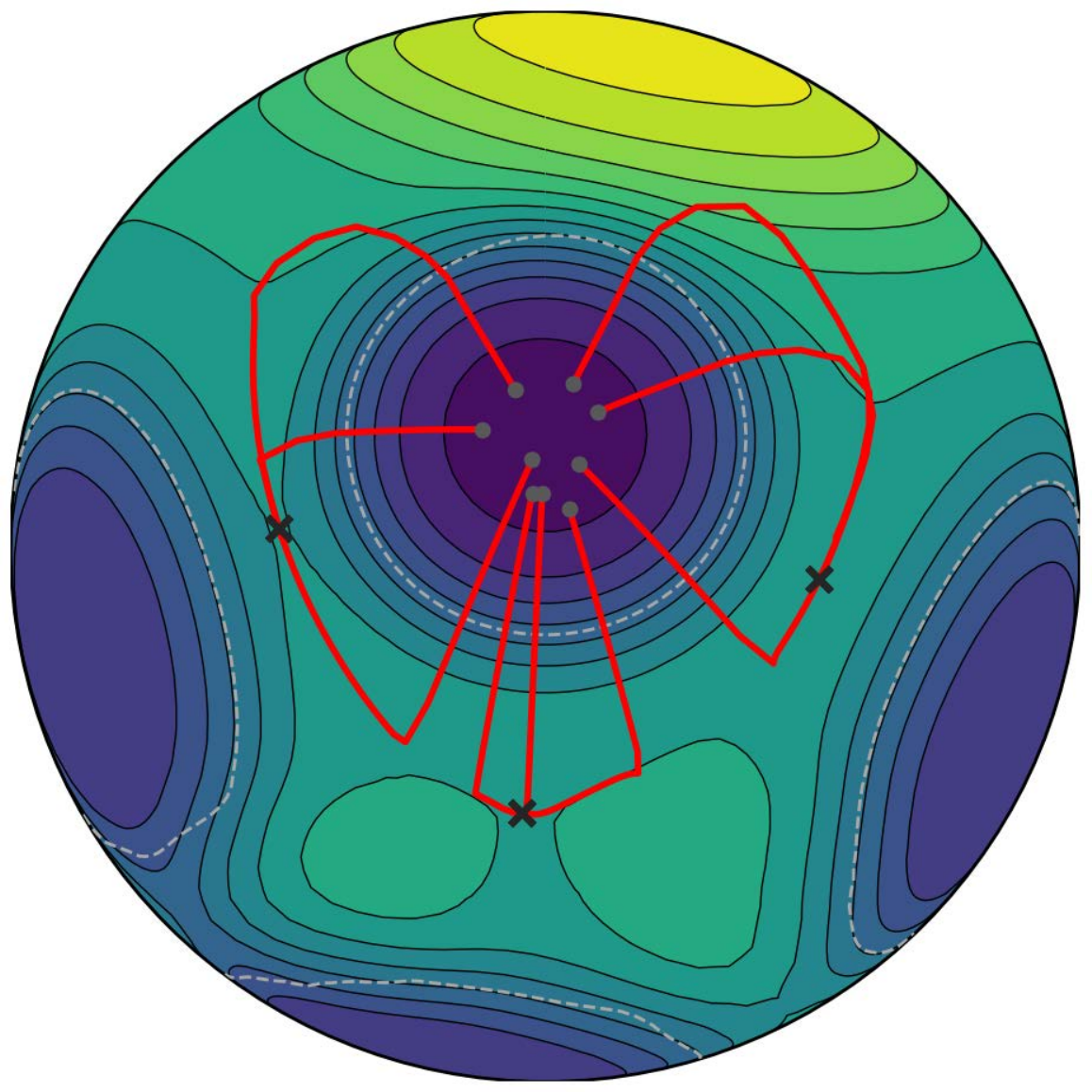}
\caption{
An illustration of a method for climbing up the energy surface from an initial state minimum to a first order saddle point. The system contains only a single spin (see \cite{SupportingInfo}) and because only the orientation of the magnetic vector can change, the energy surface can be mapped onto a sphere.  The local energy minima (blue) are separated by an energy ridge where points of low energy correspond to first order saddle points (black $\times$). The red curves illustrate saddle point search paths starting from various points. 
In this illustration, the system is made to follow the gradient of the energy until the lowest eigenvalue of the Hessian becomes negative 
(at the white dashed line). Beyond that point, the system is displaced along the effective force 
given by Eq.~\eqref{eq: effective force}.
This brings the system to a first order saddle point on the energy surface. 
}
	\label{fig: mmf single spin}
\end{figure}

This approach has not, to the best of our knowledge, been applied previously to a magnetic system. 
Here, the configuration space $\mathcal{M}_{\mathrm{phys}}$ 
is given by the direct product of $N$ spheres.
In order to apply the mode following method, knowledge of second order derivatives is required
but, as is well-known from the theory of Riemannian manifolds,  
they need
to be treated with special care.
The Hessian
can be calculated by application of covariant derivatives, but
their evaluation in spherical coordinates is usually cumbersome and suffers from singularities at the poles.

A more convenient approach 
is offered by viewing the configuration space as being embedded in a surrounding euclidean space $\mathcal{E}\supset\mathcal{M}_{\mathrm{phys}} $. In this larger space, second order derivatives are easily performed. The Hessian in the physical subspace can then be reconstructed by a projection operator approach~\cite{absil_extrinsic_2013}. For any scalar function $f$ on the manifold $\mathcal{M}_{\mathrm{phys}} $, this true Hessian is defined as
\begin{equation}
	\mathrm{Hess}~f(\vecext{x})[\vecext{z}] = P_{\vecext{x}} \partial^2 \bar{f}(\vecext{x}) \vecext{z} + W_{\vecext{x}} ({\vecext{z}}, P_{\vecext{x}}^\perp \partial \bar{f}),
    \label{eq: hessian generalized}
\end{equation}
where $z$ is a vector tangent to the manifold, $P_x$ and $P_x^\perp$ are the projectors onto the tangent space and onto the normal space, respectively, to the surface of the manifold at a point $x$ on the sphere, $\bar{f}$ is the smooth extension of $f$ to the Euclidean space, and $W_x$ is the Weingarten map at $x$. 
The Weingarten map, sometimes also referred to as the shape operator, describes the curvature of 2D surfaces in terms of an embedding space.

For a Hamiltonian $\mathcal{H}$ of a spin system, the matrix representation of the Hessian is obtained by its action on the basis vectors
(see \cite{SupportingInfo}), i.e.,
\begin{equation}
	\hess_{ij} = T_i^T \carthess_{ij} T_j - T_i^T I (\vec{n}^j\cdot\nabla^j\bar{\mathcal{H}}) T_j,
    \label{eq: hessian final}
\end{equation}
where the indices $i$ and $j$ denote spins, $\carthess=\partial^2 \bar{\mathcal{H}}(x)$, $I$ is the $3\times3$ unit matrix and $T_i$ is the $3\times2$ matrix that transforms into the tangent space of spin $i$.
%
As the Hessian matrix given by Eq. \eqref{eq: hessian final} is represented in $2N$, 
the evaluation of an eigenmode in the $3N$-representation
requires a
transformation, i.e. $\vecext{\lambda}|^{3N} = T\vecext{\lambda}|^{2N}$ (see \cite{SupportingInfo}).
%
%

\begin{figure}[!h]
    \includegraphics[width=0.95\linewidth]{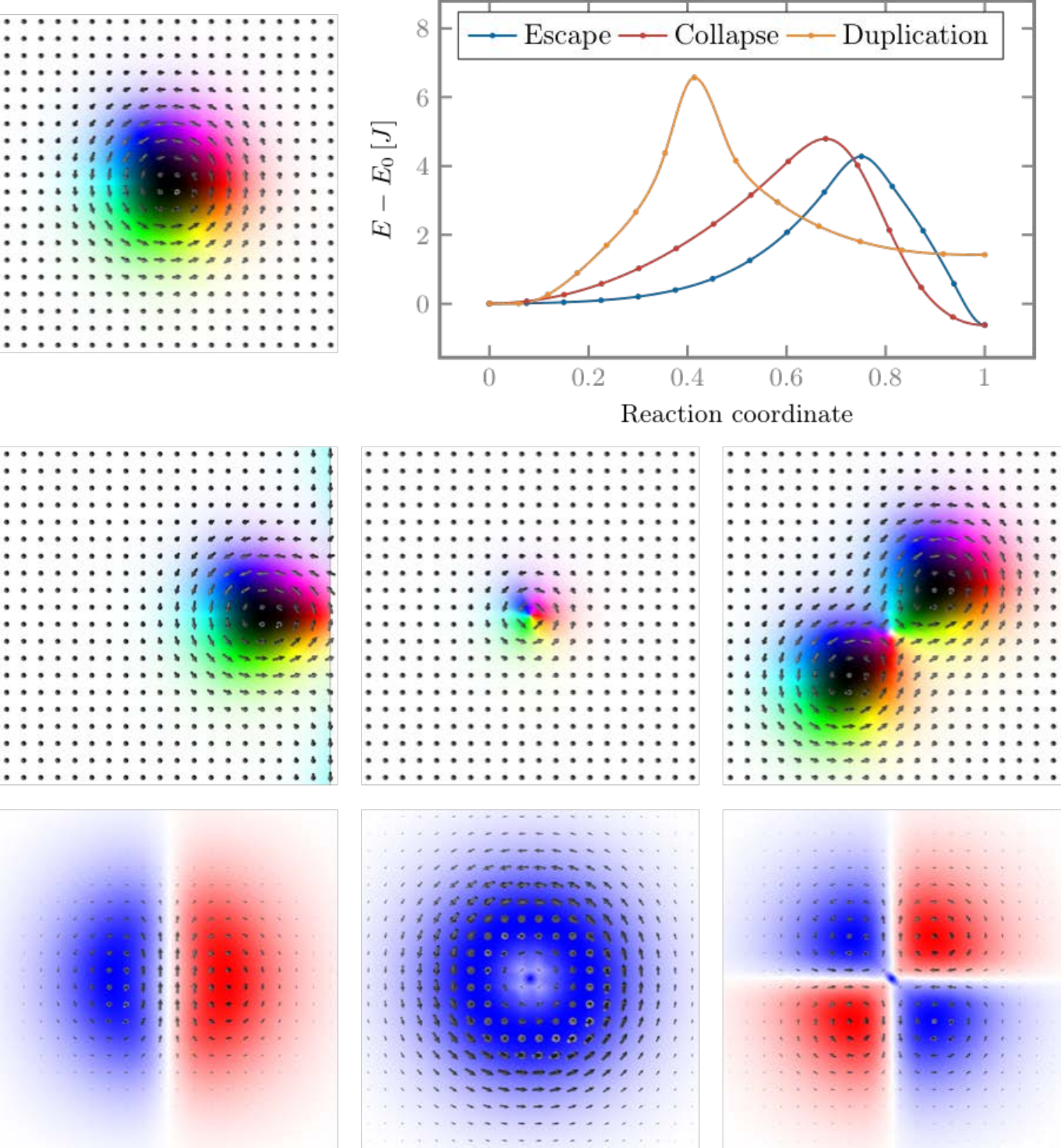}
    \caption{
    Top panel: (Left) Minimum energy configuration of the skyrmion.
               (Right) Minimum energy paths for the three types of transitions found: Duplication, collapse and escape. The reaction coordinate is the scaled total displacement along the path. The energy is given in units of the exchange coupling constant, $J$.
    Middle panel: Saddle point configurations found for escape through a boundary (left), radial collapse (middle) and duplication (right). 
    Bottom panel: The translational, breathing and elliptical eigenmodes of the skyrmion in the minimum energy configuration, each leading to the saddle point displayed directly above.
    }
    \label{fig: gneb}
\end{figure}

The formulation of the constrained Hessian given by Eq.~\eqref{eq: hessian final} and the saddle point search method described above have been implemented in the \textit{Spirit}~\cite{spirit} software and used here to analyse transitions from a magnetic skyrmion state.  
%
The energy of the system is described by an extended Heisenberg model
\begin{equation}
	\mathcal{H} = -\mu_\mathrm{S} \sum\limits_{i=1}^{N
	} \vec{H}\cdot\vec{n}_i - J \sum\limits_{\braket{ij}}\,  \vec{n}_i\cdot\vec{n}_j - D \sum\limits_{\braket{ij}} \hat{\vec{d}}_{ij} \cdot (\vec{n}_i\times\vec{n}_j),
	\label{eq: Hamiltonian}
\end{equation}
where $\vec{H}$ is a uniform external magnetic field, $\vec{n}_i$ is the magnetic moment of 
spin $i$, $J$ is the exchange coupling between nearest neighbor spins, 
the parameter $D$ 
and unit vector $\hat{\vec{d}}_{ij}$ give the Dzyaloshinskii-Moriya vector in the plane of the lattice parallel to the vector connecting the two nearest neighbors $i$ and $j$.
The sums include only distinct nearest neighbor pairs. 
The system consists of 40$\times$40 spins on a square lattice with free boundary conditions and
the parameter values 
and field strength
are chosen to be the same as in a recent
theoretical study \cite{Rybakov_15} (values are given in \cite{SupportingInfo}).
The Bloch skyrmion, shown in Fig.~\ref{fig: gneb}, is metastable with respect to the ferromagnetic phase.


Fig.~\ref{fig: gneb} shows an illustration of the eigenvectors corresponding to the three lowest eigenvalues. They correspond to translation, breathing and elliptical distortion
 \cite{lin_internal_2014}. 
In order to escape from the convex region, we have chosen here to follow each one of these three modes until the corresponding eigenvalue becomes negative. After that, the system is displaced in the direction of the effective force given by 
Eq.~\eqref{eq: effective force}, where $\hat{\vecext{\lambda}}$ is the eigenvector of the initially selected mode rather than the one with lowest eigenvalue.

\begin{figure*}[!hbt]
      \includegraphics[width=1\linewidth]{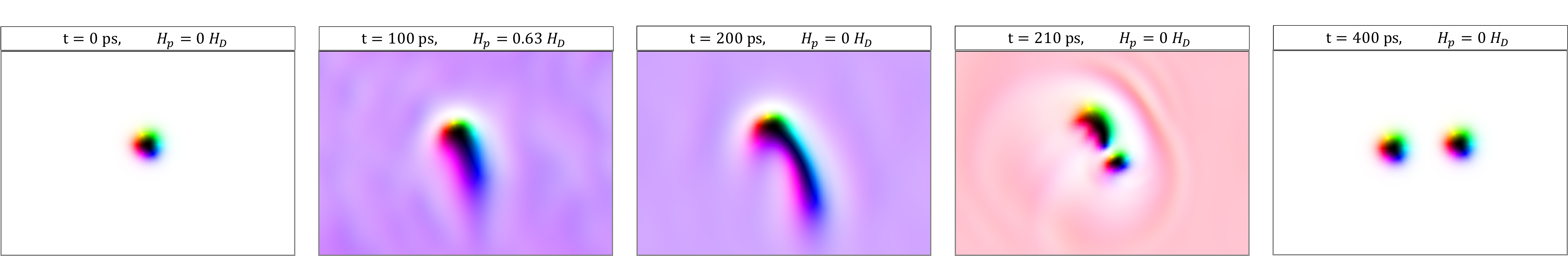}
    \caption{
    	Snapshots from a dynamical simulation where the duplication of a skyrmion is induced by 
	applying a magnetic pulse over 200 psec, giving a total field that is tilted with respect to the normal to the plane. 
	(see \cite{SupportingInfo} for parameter values). The labels on top of the frames give the time and magnetic field of the
	pulse (which adds to a constant field in the direction of the normal to the plane).
	The pulse is applied at time t=0 and lasts until t=200 ps.
At the end of the pulse, an elongated defect is formed which later splits up (at t=209 ps) to form two skyrmions.
}
    \label{fig: mitosis dynamics}
\end{figure*}

%
By following the translational mode, the mode that has lowest eigenvalue initially, the skyrmion moves towards the edge of the system and eventually, as it is pushed along the effective force, converges on a saddle point corresponding to an escape through the edge. Relaxation of the system after a slight displacement along the unstable mode at the saddle point brings the system to the ferromagnetic state.
Similarly, by following the breathing mode, the second lowest eigenvalue mode, the skyrmion shrinks and eventually, 
converges to a saddle point corresponding to collapse. Again the final state of the transition is the ferromagnetic state.  
However, by following the elliptical mode, the skyrmion becomes stretched and 
converges on a saddle point corresponding to a division of the skyrmion and formation of two skyrmions. While the initial state contains one skyrmion, the final state contains two. 

The energy along the minimum energy paths for the three transitions is shown in Fig.~\ref{fig: gneb}. These were calculated using the geodesic nudged elastic band method \cite{bessarab_method_2015} using an initial path formed by linear interpolation between the intial state and the saddle point, as well as between the saddle point and the final state. The activation energy for the collapse and escape is quite similar, but the activation energy for the duplication is higher 
in this case.
The relative height of the energy barriers for the three processes depends on the parameter values. 
A calculation using a 10\% smaller field and 60\% larger value of $D/J$ gives a lower activation energy for duplication than the other two transitions \cite{SupportingInfo}.

The shape of the three curves is quite different. The total displacement along the paths
from the initial state to the saddle point is shortest for the duplication, while the displacement of the skyrmion to the boundary and to the saddle point for escape involves the largest displacement. The energy profile for the collapse is similar to what has been presented before, a gradual shrinkage of the skyrmion to the saddle point followed by abrupt drop of the energy to that of the ferromagnetic state  \cite{bessarab_method_2015,Lobanov16}.

The duplication of the skyrmion leads to an increase in the energy of the system because the skyrmion is only metastable with respect to the ferromagnetic state for this set of parameters, while the other two transitions lead to a decrease in the energy.
The energy along the minimum energy path past the saddle point for duplication contains important information about skyrmion interaction. It shows how the repulsive interaction between skyrmions varies with the distance between them. 
The minimum energy path going from right to left along the mininum energy path in Fig. 2 shows how two skyrmions can merge to form one skyrmion.
%

The duplication transition identified here has not been described previously, but 
may turn out to be an important mechanism for generating skyrmions.
The first order saddle point on the energy surface corresponding to this mechanism can be found for a wide range of parameter
values, as will be described in detail in a later publication.
For example, we have found the duplication saddle point for a Pd/Fe bilayer on an Ir(111) substrate, a system that has been studied
extensively, using parameter values that give close agreement with experimental results \cite{bessarab_2018,Malottki_2018}.
There, the energy barrier for duplication turns out to be slightly lower than that of collapse, 78 meV vs. 80 meV.

The question now arises whether it is possible to induce the duplication event dynamically.
We address this by performing a dynamical simulation where a short magnetic pulse is applied in addition to the stationary magnetic
field. 
The pulse is represented by a uniform magnetic field which has both in-plane and out-of-plane components. 
For the duration of the pulse, 200 ps, the total magnetic field is tilted (see \cite{SupportingInfo}).  
This is known to be an efficient way of exciting nonlinear skyrmion dynamics \cite{Heo16}.
The saddle point searches show that an excitation of the elliptical mode can induce duplication. 
Fig.~\ref{fig: mitosis dynamics} shows shapshots from the simulation.
During the pulse, an elongated structure is formed.  After the pulse has been turned off, it splits up and eventually two skyrmions remain in the system.
In addition to the elliptic elongation, the duplication also requires the deformed structure to be bent in order for it to divide up. The bending corresponds to another eigenmode of the elongated skyrmion.
%
More efficient and/or reliable ways of achieving the duplication process could be devised, for example by pinning the skyrmion, introducing defects or by other more elaborate external stimuli. 
The simulation described here is merely meant to show that it is possible to induce skyrmion duplication in a rather simple way.



We have presented here a method that can be used to search for transition mechanisms and determine the activation energy for transitions in magnetic systems. It represents an adaptation of mode following methods that have been used for several years in studies of atomic rearrangements. The fact that magnetic transitions involve rotations of the magnetic vectors leads to constraints that need to be incorporated into the Hessian matrix in order to determine the low lying eigenvalues and corresponding eigenvectors. 
Saddle point searches where only the initial state of the transition is specified, such as the ones carried out here, are more challenging than a minimum energy path calculation where both the initial and final states are specified, 
but they have the advantage of being able to reveal unexpected transitions and unknown final states.
In the application calculations presented here, we have chosen to drive the system out of the convex region by following a particular
mode.
We note that, eigenvalues may cross and modes change direction 
as the system is pushed along a specific mode, so care should be taken to remain on the same mode throughout the saddle point search
(see \cite{SupportingInfo}). 

%

Applications of the saddle point search method presented here to three-dimensional systems can be expected to yield an even larger variety of mechanisms than for two-dimensional systems since more possibilities open up with the additional dimension. For example, cylindrical skyrmions can contract and form bobbers \cite{Rybakov_15} and possibly other previously unknown magnetic structures.

A saddle point search method that only requires the initial state as input 
(unlike the GNEB method where the final state also needs to be specified),
can be used to sample an energy surface in a systematic way and search for possible states of the system, each state corresponding to a local energy minimum.
The search takes the system from one local minimum to another via first order saddle points. This can be used as the basis of a simulation of the long time scale evolution of a system that undergoes thermally activated transitions that may be enhanced by an external field.
For each state visited, multiple saddle point searches are carried out to identify the most relevant, low energy saddle points. The transition rate corresponding to each saddle point can be estimated using harmonic transition state theory, and a random number used to choose which transition will occur next based on the normalized rates. This is referred to as adaptive kinetic Monte Carlo algorithm \cite{Henkelman_2001}. The simulated time can be estimated from the sum of rates.
This approach has been used as the basis for simulations of long time scale evolution of complex atomic systems where atomic rearrangement events can involve non-intuitive displacements of multiple atoms \cite{Chill14,Pedersen09}. The adaptation presented here of the saddle point search method to spins opens the possibility of carrying out such long time scale simulations of magnetic systems.


\begin{acknowledgments}
	This work was funded by Icelandic Research Fund (grants 185405-051 \& 184949-051),
Academy of Finland (grant 278260), Russian Foundation of Basic Research (grant RFBR 18-02-00267 A),
EU-H2020 MAGicSky (grant 665095), DARPA TEE 
(\#HR0011831554) from DOI, and DFG CRC 1238 (project C01). 
\end{acknowledgments}



\clearpage
\pagebreak
\onecolumngrid

\setcounter{equation}{0}
\setcounter{figure}{0}
\renewcommand{\thesection}{S\arabic{section}.}
\renewcommand{\thefigure}{S\arabic{section}.\arabic{figure}}
\renewcommand{\theequation}{S\arabic{section}.\arabic{equation}}

{\centerline {\bf  \large Supplemental Material for ``Duplication, collapse and escape of }}
{\centerline {\bf \large magnetic skyrmions revealed using a systematic saddle point search method''}}


\vskip 0.3 true cm

\centerline {Gideon P. M\"uller,$^{1,2,3}$ Pavel F. Bessarab,$^{1,4}$ Sergei M. Vlasov,$^{1,4}$ Fabian Lux$^{2,3}$}
\centerline {Nikolai S. Kiselev,$^{2}$ Stefan Bl\"ugel,$^{2,3}$ Valery M. Uzdin,$^{4,5}$ and Hannes J\'onsson$^{1,6}$}

\vskip 0.1 true cm
\centerline {\it $^1$Science Institute and Faculty of Physical Sciences, University of Iceland, VR-III, 107 Reykjav\'{i}k, Iceland}
\centerline {\it $^2$Peter Gr\"unberg Institut and Institute for Advanced Simulation, Forschungszentrum J\"ulich and JARA, 52425 J\"ulich, Germany}
\centerline {\it $^3$RWTH Aachen University, D-52056 Aachen, Germany}
\centerline {\it $^4$ITMO University, 197101, St. Petersburg, Russia}
\centerline {\it $^5$Dpt.~of Physics, St. Petersburg State University, St. Petersburg 198504, Russia}
\centerline {\it $^6$Dpt.~of Applied Physics, Aalto University, FIN-00076 Espoo, Finland}


\ 
\vskip 0.5 true cm

\stepcounter{section}
\section{Evaluation of the Hessian for spin systems}
Eigenvectors of the Hessian matrix are needed in the saddle point search method.
However, 
the second derivatives do not have a direct
geometrical meaning and
need to be treated within the theory of Riemannian manifolds
~\cite{Nakahara2003}.
Below, a representation is derived for the Hessian of spin systems where the spin length is fixed,
i.e where the manifold $\mathcal{M}_\mathrm{phys}$ of physical states is composed of the direct product of $N$ spheres
\begin{equation}
	\mathcal{M}_\mathrm{phys} = \bigotimes\limits_{i=1}^{N} S^2 \subset \mathbb{R}^{3N}.
\end{equation}
Therefore, $\mathcal{M}_\mathrm{phys}$ is a submanifold of the embedding euclidean space $\mathcal{E} = \mathbb{R}^{3N}$.

Due to the singularities that can arise when using spherical coordinates, we choose to use a $3N$ cartesian representation of the spin coordinates. 
%
As the Hamiltonian $\mathcal{H}$ can be extended to a function $\bar{\mathcal{H}}$ which is defined on $\mathcal{E}$, the derivatives can be calculated easily.
We denote the gradient taken in the embedding space $\mathcal{E}$ as $\partial\bar{\mathcal{H}}$ and the gradient taken on the manifold  $\mathcal{M}_\mathrm{phys}$ as $P_x\partial\bar{\mathcal{H}}$, i.e., the projection of the gradient onto the tangent space.
The matrix of second derivatives in the embedding space is denoted $\partial^2\bar{\mathcal{H}}$.

In order to calculate the second derivatives on $\mathcal{M}_\mathrm{phys}$ while keeping the description in Cartesian coordinates, the projector approach described in Ref.~\cite{absil_extrinsic_2013Supp} is used.
For any scalar function $f$ on the manifold $\mathcal{M}_{\mathrm{phys}} $, this true, covariant Hessian is defined as
\begin{equation}
	\mathrm{Hess}~f(\vecext{x})[\vecext{z}]  = P_{\vecext{x}} \partial^2 \bar{f}(\vecext{x}) \vecext{z} + W_{\vecext{x}} ({\vecext{z}}, P_{\vecext{x}}^\perp \partial \bar{f}).
    \label{eq: supplement hessian generalized}
\end{equation}
%
Here, $W_x$ denotes the Weingarten map which for any vector $v$ at a point $x$ of a spherical manifold is given by
\begin{equation}
	W_x(z,v) = -z x^T v,
\end{equation}
where $z$ is a tangent vector to the sphere at $x$.
In order to calculate the Hessian, $v=P_x^\perp \partial \bar{\mathcal{H}}$ is inserted to give
\begin{equation}
    W_x (z, P_x^\perp \partial \bar{\mathcal{H}})
      = -z x^T P_x^\perp \partial  \bar{\mathcal{H}}
      = -z x^T x x^T \partial  \bar{\mathcal{H}}
      = -z x^T \partial  \bar{\mathcal{H}},
\end{equation}
where $x^T \partial  \bar{\mathcal{H}}$ is the scalar product of the spin with the gradient. 

In \textit{Spirit}~\cite{spiritSupp}, the Hessian is implemented in matrix representation, so we switch notation and drop the subscript $x$.
For spin indices $i$ and $j$, we write the gradient $\partial \bar{\mathcal{H}}$ as $\nabla_i\bar{\mathcal{H}}$ (now to be understood as a three-dimensional object) and the second derivative $\partial^2\bar{\mathcal{H}}$ as $\bar{H}$.
We reformulate Eq.~\eqref{eq: supplement hessian generalized} within the Euclidean representation as a $3N\times3N$ matrix 
\begin{equation}
H|^{3N} = (H_{ij}|^{3N} ) = 
\begin{pmatrix}
H_{11}|^{3N}  & H_{12}|^{3N} & \cdots \\
H_{21}|^{3N} & H_{22}|^{3N} & \cdots \\
\vdots & \vdots & \ddots
\end{pmatrix}
\end{equation}
which consists of $N^2$ blocks that correspond to the different spin-spin subspaces. This matrix representation  of the Hessian is obtained by acting with Eq.~\eqref{eq: supplement hessian generalized} on the euclidean basis vectors of the extended space. Here, each of the $3\times 3$ subspace matrices is defined as
\begin{equation}
	H_{ij}|^{3N} = P_i \bar{H}_{ij} - \delta_{ij} I \vec{n}_j\cdot\nabla_j\bar{\mathcal{H}},
\end{equation}
where $I$ denotes the $3\times3$ unit matrix and $\vec{n}_j$ the normalized direction of spin $j$. 
The resulting matrix $H|^{3N}$ will, however, have $N$ eigenvectors orthogonal to the tangent space of $\mathcal{M}_\mathrm{phys}$, representing unphysical degrees of freedom in the embedding space $\mathcal{E}$.
In order to remove these unphysical degrees of freedom, 
the matrix is transformed using the tangent basis to $\mathcal{M}_\mathrm{phys}$ and written as
$H_{ij} = T_i^T H_{ij}|^{3N} T_j$, where $T_i$ is the basis transformation matrix of spin $i$ fulfilling $T^TP = T^T$ and $T^TT = I|^{2N}$.
Thereby, the true Hessian of Eq.~\eqref{eq: supplement hessian generalized}, in the $2N\times2N$ matrix representation, becomes $H = (H_{ij})$ with the spin-spin blocks defined as
\begin{equation}
	H_{ij} = T_i^T \bar{H}_{ij} T_j - T_i^T I (\vec{n}_j\cdot\nabla_j\bar{\mathcal{H}}) T_j,
\end{equation}
and which now contains only the physical degrees of freedom.
From this matrix, the eigenmodes $\lambda|^{2N}$ are calculated. The $3N$ representation is obtained as $\lambda|^{3N} = T\lambda|^{2N}$.

%
The transformation matrix $T_i$ can be a $3\times2$ matrix of two tangent vectors to spin $i$, that can be obtained from any random vector and another vector found by orthogonalization.
Here, we have chosen the unit vectors of spherical coordinates $\theta$ and $\varphi$, which gives
%
\begin{equation}
	T = \{\vec{e}_\theta, \vec{e}_\varphi\}
      = \begin{pmatrix}
          \cos\theta\cos\varphi & -\sin\varphi \\           
          \cos\theta\sin\varphi &  \cos\varphi \\
          -\sin\theta & 0
        \end{pmatrix} \\
      = \begin{pmatrix}
          zx/r_{xy} & -y/r_{xy} \\           
          zy/r_{xy} &  x/r_{xy} \\
           - r_{xy} & 0
        \end{pmatrix}
\end{equation}
where $r_{xy} = \sin\theta = \sqrt{1-z^2}$.
Note that the poles need to be excluded, but one may simply choose $\vec{e}_x$ and $\vec{e}_y$ and orthogonalize them with respect to the spin vector.
%

%
Finally, the Hessian matrix in the embedding space $\mathcal{E} = R^{3N}$ is needed, denoted $H_{ij}|^{3N}$. 
For a Hamiltonian of the form
\begin{equation}
	\begin{alignedat}{2}
	\mathcal{H} =
      - \mu_\mathrm{S}\, \sum\limits_{i=1}^{N} \vec{H}\cdot\vec{n}_i
     - \sum\limits_{i=1}^{N} K_i (\hat{\vec{K}}\cdot\vec{n}_i)^2
     - \sum\limits_{\braket{ij}}\, J_{ij} ~ \vec{n}_i\cdot\vec{n}_j
      -  \sum\limits_{\braket{ij}}\, \vec{D}_{ij} \cdot (\vec{n}_i\times\vec{n}_j),
	\end{alignedat}
\end{equation}
where $\vec{H}$ is the external magnetic field, $K$ are uniaxial anisotropies 
(set to zero in the calculations presented here), $J_{ij}$ are the exchange constants
and $\vec{D}_{ij}$ the Dzyaloshinskii-Moriya vectors of the unique interaction pairs denoted by $\braket{ij}$,
the spin-spin matrix blocks of $\bar{H}=\partial^2\bar{\mathcal{H}}$ become

\vskip 0.2 true cm

\textit{diagonal blocks}:
\begin{equation}
	\bar{H}_{i=j} = -2K_i
    \begin{pmatrix}
       K_xK_x & K_xK_y & K_xK_z \\[0.5em]
       K_yK_x & K_yK_y & K_yK_z \\[0.5em]
       K_zK_x & K_zK_y & K_zK_z
    \end{pmatrix}
\end{equation}

\textit{off diagonal blocks}:
\begin{equation}
  \bar{H}_{i\neq j} =
    \begin{pmatrix}
      -(J_{ij}+J_{ji})    & (-\vec{D}_{ij}+\vec{D}_{ji})_z  & -(-\vec{D}_{ij}+\vec{D}_{ji})_y \\[0.5em]
      -(-\vec{D}_{ij}+\vec{D}_{ji})_z & -J_{ij}-J_{ji}      & (-\vec{D}_{ij}+\vec{D}_{ji})_x \\[0.5em]
      (-\vec{D}_{ij}+\vec{D}_{ji})_y  & -(-\vec{D}_{ij}+\vec{D}_{ji})_x & -J_{ij}-J_{ji}
    \end{pmatrix}.
\end{equation}
%


\vskip 0.5 true cm

\stepcounter{section}
\section{Saddle point search method}
%
%
The saddle point search method is an iterative process where the spin configuration is modified according to
a force acting on the spins.
Each iteration involves finding the Hessian and its eigenvalues and eigenvectors.
An important aspect of the adaptation of this method to spin systems is the evaluation of the Hessian for 
the curved configuration space.
The optimization method involves rotating the spins in the direction of the force using the velocity projection method described in \cite{bessarab_method_2015Supp} until the force is zero.
Outside the convex region, where one or more of the eigenvalues of the Hessian is negative, 
the effective force, $\vec{F}^\mathrm{eff}$, is given by Eq. (4) in the main text.

There are various approaches for escaping from the convex region around the initial state minimum:
\begin{itemize}
\item Follow a (constant) random direction.
\item Make an initial (small) random displacement and then follow the gradient.
\item Pick one of the eigenvectors of the Hessian and follow it. 
\end{itemize}
In the illustrative test problem shown in Fig. 1 in the main text, the second choice was made (due to
the reduced dimensionality of the configuration space). In the saddle point searches for the skyrmion the third choice was made.

The iterative optimization using  $\vec{F}^\mathrm{eff}$ may lead back into the convex region.
Also, eigenvalues may become degenerate and the lowest eigenvalue mode may change significantly throughout a calculation. When following a given mode, the scalar product of the current and previous mode is evaluated and, if a significant change in direction has occurred, the search is carried out for the mode that has the largest scalar product with the previous mode. This approach works due to the fact that eigenmodes are mutually orthogonal and should change continuously throughout 
configuration space.

In special cases the gradient may become orthogonal to the mode to within a given tolerance.
This case needs to be treated specially.
The following algorithm was used in the present studies:
%
\begin{itemize}
  \item When the mode is not orthogonal to the gradient
    \begin{itemize}
    \item if the mode has a negative eigenvalue: follow $\vec{F}^\mathrm{eff}$
  	\item if the mode has a zero or positive eigenvalue: 
                  follow the mode.
    \end{itemize}
  \item When the mode is orthogonal to the gradient
    \begin{itemize}
    \item if the eigenvalue is zero (to within a tolerance): follow the gradient
    \item if the eigenvalue is positive: follow $\vec{F}^\mathrm{eff}$.
    \end{itemize}
\end{itemize}
%


\vskip 0.5 true cm

\stepcounter{section}
\section{Parameter values}
\noindent{\bf 1. \ For single spin illustration}
\vskip 0.3 true cm

The energy surface of the single-spin system shown in Fig. 1 in the main text to illustrate the saddle point search method 
is a sum of 7 Gaussians of the form
\begin{equation}
	H = \sum\limits_i H_i  = \sum\limits_i a_i \exp\left(  -\frac{l_i^2(\vec{n})}{2\sigma_i^2} \right),
\end{equation}
with parameters given in Table 1.

\begin{table}[htb]
	\caption{Parameters of the Gaussians in the energy surface of the single-spin system shown in Fig. 1 in the main text.
	}
	\begin{filecontents*}{param_gaussian.txt}
-1.1   0.06  -0.2   0.0  -0.9
 0.8   0.15  -1.0   0.2  -0.2
-0.9   0.1    1.0  -0.2  -0.1
 0.09  0.03   0.8   0.5  -0.8
 0.13  0.03   0.8  -0.5  -0.7
-0.9   0.1    0.5   1.2  -0.4
-0.9   0.1    0.2  -0.9  -0.4
\end{filecontents*}
\pgfplotstabletypeset
		[
			header=false,
			display column names={a,$\sigma$,$p_x$,$p_y$,$p_z$},
			every head row/.style={before row=\toprule, after row=\midrule},
			every last row/.style={after row=\bottomrule}
		]
		{param_gaussian.txt}
	\label{Table: MMF Single Spin}
\end{table}

\vskip 0.5 true cm


\noindent{\bf 2. \ For Hamiltonian in the skyrmion calculations}
\vskip 0.3 true cm

The values of the parameters in the Heisenberg type Hamiltonian, Eq. (4) in the main text, 
were chosen to be the same as in a previous theoretical study \cite{Rybakov_15Supp},
$D=0.45~J$ and $H= 0.8~H_\mathrm{D}$,
where $H_\mathrm{D} = D^2/(\mu_\mathrm{S} J)$ is the critical field strength. For the parameter values used here
and setting J=1$~\mathrm{meV}$, gives
$H_\mathrm{D}$ = 3.5$~\mathrm{T}$.

\begin{figure}[!t]
	\includegraphics[width=0.5\linewidth]{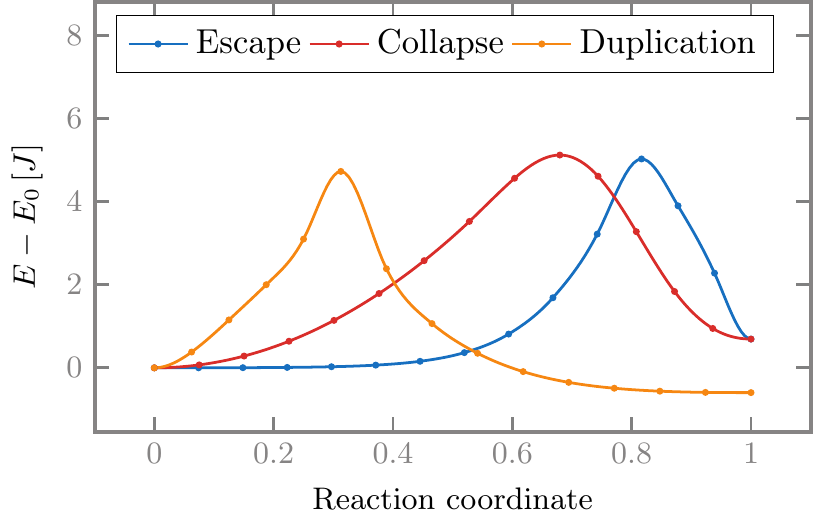}
    \caption{
    Minimum energy paths for the three types of transitions 
    using parameter values of 
       $\ H=0.71~H_\mathrm{D}, \ D=0.7~J$.
    The reaction coordinate is the total rotation of all spins, normalized by the full length of the path.
    }
    \label{fig: supplement gneb}
\end{figure}

%


\vskip 0.5 true cm

\section{Minimum energy paths}
In addition to the minimum energy paths shown in Fig. 2 in the main text, 
a calculation was carried out with another set of parameters, 
$D=0.7~J$ and $H=0.71~H_\mathrm{D}$.
%
In this case, the activation enegy is quite similar for the three transitions, the duplication barrier being lowest, as can be seen from Fig.~\ref{fig: supplement gneb} in this Supplemental Material.
Here, the skyrmion state is lower in energy than the ferromagnetic state. 

\vskip 0.5 true cm

\section{Dynamical simulation}

In the dynamical simulation of the duplication process, a magnetic pulse is applied over a time period of 200 ps.
The pulse has a magnetic field directed along the vector \{x,z\}=\{0.8,-0.61\} while
the static magnetic field is directed along the normal along the 
vector \{x,z\}=\{0,1\} so the pulse partly cancels out the static field and leads to a tilt.
The static field has magnitude $H=0.8~H_\mathrm{D}$ but during the pulse the total magnetic field drops to $H_p=0.63~H_\mathrm{D}$.   $D=0.45~J$ as in the 
calculations shown in Fig. 2 in the main text. The time evolution was carried out using the semi-implicit algorithm of Mentink
{\it et al.} \cite{Mentink10} with a time step of 0.01~ps.




\begin{thebibliography}{99}
\bibitem{FirstOrderSP}{
At a first order saddle point the gradient of the energy vanishes and the Hessian, 
the matrix of second derivatives of the energy, has one and only one negative eigenvalue.
}
\end{thebibliography}
\end{document}